\documentclass[referee]{raa}

\usepackage{graphicx,times}
\usepackage{natbib}
\usepackage{multirow}
\usepackage{amssymb,amsmath}
\bibpunct{(}{)}{;}{a}{}{,}
\graphicspath{{./}{Figures/}}

\usepackage[pagebackref=true]{hyperref}
\usepackage{color}
\usepackage{graphicx} 

\title{The Velocity Aberration Effect of the CSST Main Survey Camera
\footnote{This research has been generously supported by the National Natural Science Foundation of China (Grant Nos. 12073047 and 12273077) and the National Key Research and Development (Grant No. 2022YFF0711500).}}

\author{Hui-Mei Feng 
      \inst{1,2,3}
   \and Zi-Huang Cao
      \inst{3,4}
   \and Man I Lam 
      \inst{3}
   \and Ran Li
      \inst{3,2,4}
    \and Hao Tian
      \inst{3}
    \and Xin Zhang
      \inst{3}
    \and Peng Wei
      \inst{3}
    \and Xin-Feng Li 
      \inst{5}
    \and Wei Wang
      \inst{6}
    \and Hugh R. A. Jones
      \inst{7}
    \and Mao-Yuan Liu
      \inst{1}
    \and Chao Liu
      \inst{3,4}
}

   \institute{Key Laboratory of Cosmic Rays (Tibet University), Ministry of Education, Lhasa 850000, China\\
        \and
            Institute for Frontiers in Astronomy and Astrophysics, Beijing Normal University, Beijing 102206, China
        \and
            National Astronomical Observatories, Chinese Academy of Sciences, Beijing 100012, China\\
        \and
            University of Chinese Academy of Sciences, Beijing 100049, China\\
        \and
            Technology and Engineering Center for Space Utilization, Chinese Academy of Sciences, Beijing 100094, China\\
        \and
            Changchun Institute of Optics, Fine Mechanics and Physics, Chinese Academy of Sciences, Changchun 130033, China\\
        \and
            Centre for Astrophysics, University of Hertfordshire, Hatfield, UK\\
    {\it zhcao@nao.cas.cn}\\
\vs\no
   {\small Received 2023 month day; accepted 202x month day}}

\begin{document}

\abstract{
In this study, we conducted simulations to find the geometric aberrations expected for images taken by the Main Survey Camera (MSC) of the Chinese Space Station Telescope (CSST) due to its motion. As anticipated by previous work, our findings indicate that the geometric distortion of light impacts the focal plane's apparent scale, with a more pronounced influence as the size of the focal plane increases. Our models suggest that the effect consistently influences the pixel scale in both the vertical and parallel directions. The apparent scale variation follows a sinusoidal distribution throughout one orbit period. Simulations reveal that the effect is particularly pronounced in the center of the Galaxy and gradually diminishes along the direction of ecliptic latitude. At low ecliptic latitudes, the total aberration leads to about 0.94 pixels offset (a 20-minute exposure) and 0.26 pixels offset (a 300-second exposure) at the edge of the field of view, respectively. Appropriate processings for the geometric effect during the CSST pre- and post-observation phases are presented.
}

\authorrunning{}            
\titlerunning{}             
\maketitle

\section{Introduction}
\label{sect:intro}

High-resolution observations from space have opened a crucial window for studying celestial objects in astrophysics, astrometry, and dynamic astronomy. Compared to ground-based observations, the velocity aberration effect is more significant for high-resolution observations in space. The combined velocities of the space telescope and Earth's orbital motions can lead to a deviation in the observed direction of celestial objects compared to the actual arrival direction of photons, and this effect is defined as the total velocity aberration. The total velocity aberration is a combination of diurnal and annual velocity aberrations. The diurnal velocity aberration results from the CSST's orbital motion (about 8 km/s), and the annual velocity aberration is caused by the Earth's orbital speed (about 30 km/s) around the Sun. 

In a simple observational system equipped with a single sensor oriented orthogonally to the optical axis, the velocity aberration effect becomes apparent as a change in the pixel-to-angle ratio, known as the apparent pixel scale, within the field of view (FOV) captured by the observation image. The term 'apparent pixel scale' (unit: arcsec/pixel), serves as a substitute for the terms 'apparent plate scale' or 'apparent focal panel scale' (unit: arcsec/mm), clearly and accurately describing the projection from the observed image to the celestial sphere. In our study, we define the 'pixel' as a unit of length in the focal panel, allowing us to apply this definition consistently, even in the theoretical focal plane without a sensor. The effect of velocity aberration becomes more pronounced as the focal plane size increases. This phenomenon could lead to a more noticeable elongation of a star's point spread function (PSF) and displacement of the star's position, particularly when the star closer to the focal plane periphery is observed.

In the case of a sophisticated observational system with multiple sensors deployed on a sizable focal plane, where the center of the focal plane aligns orthogonally to the optical axis, the velocity aberration effect becomes more severe and intricate. However, the optical axis within the focal plane may or may not be centered, as seen in systems like the Hubble Space Telescope (HST) or the CSST. The focal plane's substantial size amplifies the velocity aberration effect's impact, especially at its edges. Furthermore, the velocity aberration effect becomes intertwined with other factors, including the complex distortion inherent to the large focal plane, variations in the wavefront induced by changes in Adaptive Optics (AO) or Precision Image Stabilization System (PISS), and the intra-pixel effect (Wang et al., in preparation). Moreover, this effect varies across different positions in the sizable focal plane, and coupled with alterations during exposure, it can introduce a differential component of the velocity aberration. These factors pose significant challenges in accurately measuring celestial objects' shapes and positions.

To ensure both high image quality and a large field of view (FOV), the Chinese two-meter space telescope CSST adapts the Cook-type off-axis three-mirror anastigmat optical design, which is without any obstructions in the optical path to guarantee high light transmission efficiency and eliminate diffraction patterns caused by mirror supports. CSST also employs active optical technology to enhance image quality further. Combined with the above advantages, within its 1.1 square degrees FOV, the Point Spread Function (PSF) 80\% Encircled Energy Radius (REE80) is expected to be lower than 0.15". The PSF ellipticity is expected not to exceed 0.15", which can be compared to the Hubble Space Telescope (HST) (\citealt{2023ApJ...958...33F}). The MSC is the primary observation instrument of CSST, accounting for over 70\% of on-orbit observation time. This instrument comprises 30 detectors, with 18 dedicated to multi-band imaging observations and 12 for spectroscopic observations (\citealt{Zhan2021}).

This study aims to investigate the velocity aberration effect on CSST MSC observations through simulations. The findings from this research will have implications for the planning of CSST sky surveys and observation strategies. Additionally, it will contribute to enhancing the data processing pipeline for CSST MSC multiband imaging in astrometry and photometry, as well as improving spectral processing for slitless spectroscopy.
The paper is organized as follows. Section~\ref{sect:theory} reveals the theoretical explanations of velocity aberration. Section~\ref{sect:sim} introduces the simulation data and simulation method. Section~\ref{sect:result} presents the apparent pixel scale change of the MSC on the orbit by the position variations of stars according to the simulation methods. Finally, Section~\ref{sect:conclusion} summarizes our findings and suggests various mitigation methods to reduce this effect.

\section{Theoretical Explanations of Velocity Aberration Effects}
\label{sect:theory}

It is well-established that the speed of light remains constant, with changes in the gravitational field being the only factor capable of modifying it. In the theoretical framework, HST first introduced the concept of velocity aberration within space observation (\citealt{arribaseffect}). Building upon several foundations (\citealt{arribaseffect, 2023AN....34430006G, 2003PASP..115..113A, 2001A&A...375..351P}), we have derived the velocity aberration effect from a macroscopic perspective with relativity theory applied to CSST. Specifically, we have investigated the influence of aberration on the apparent pixel scale of CSST MSC. The derivation is presented as follows.

\begin{figure}
   \centering
   \includegraphics[width=0.4\textwidth, angle=0]{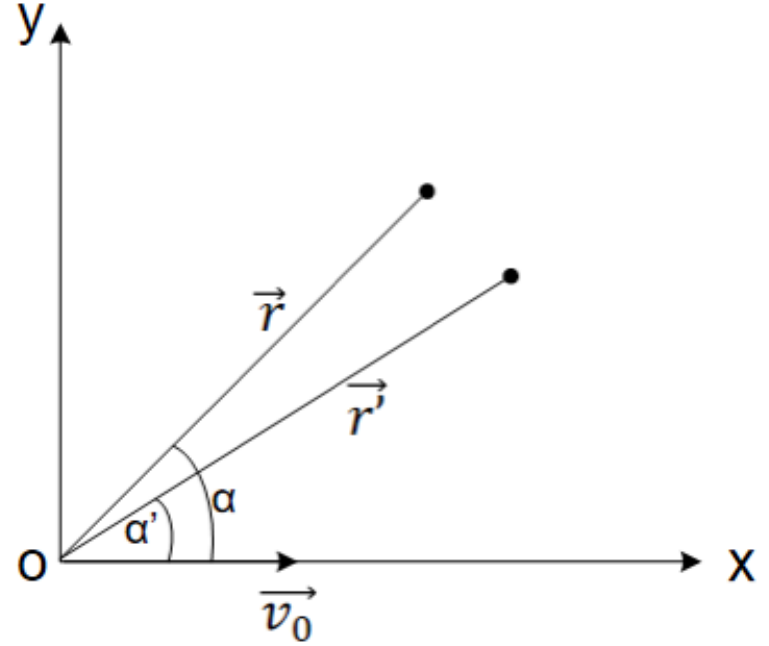}
   \caption{Relativistic interpretation of velocity aberration when $\alpha$ is an acute angle.}
   \label{Fig01}
   \end{figure}

In Figure \ref{Fig01}, $\overrightarrow{v_0}$ represents the synthesis of the combined speed of the orbital velocity of CSST and the speed of the Earth revolving around the Sun. $\overrightarrow{r}$ indicates the direction from which photons emitted by celestial bodies arrive, while $\overrightarrow{r'}$ signifies the direction in which these photons are observed. The three vectors $\overrightarrow{v_0}$, $\overrightarrow{r}$, and $\overrightarrow{r'}$ are coplanar. Suppose a scenario where a telescope records a star at point $O$ (disregarding the internal light path of the telescope) and establishes a stationary coordinate system $\Sigma$ with $O$ as the origin. The x-axis aligns with the CSST moving direction of $\overrightarrow{v_0}$, and the y-axis is perpendicular to the x-axis with the right-hand system. Both the x-axis and y-axis lie within the $\Sigma$ plane. In parallel, another coordinate system $\Sigma'$ is established. It shares its coordinate axes' orientation with those of $\Sigma$ but has a relative motion. Utilizing the Lorentz transformation, the change in the pixel scale of the focal plane caused by velocity aberration is inferred
\begin{equation}\label{eq03} 
   \frac{d \alpha'}{d \alpha} = \frac{\sqrt{1 - (\frac{v_0}{c})^2}}{1 + \frac{v_0}{c} \cos{\alpha}} = \frac{sin \alpha'}{sin \alpha}
\end{equation}
where $\alpha$ or $\alpha'$ is the angular distance between the direction of photon arrival or observation from celestial bodies relative to the motion direction of the CSST. The symbol $d$ represents a differential operator. $\frac{d \alpha'}{d \alpha}$ is the radial change in pixel scale, while $\frac{sin \alpha'}{sin \alpha}$ is the tangential change in pixel scale. Remarkably, these two values are identical, indicating that the pixel scale alteration in the focal plane within the FOV is isotropic and centrally symmetric.

The proof presented above pertains to the case when $\alpha$ is an acute angle. It can be readily demonstrated that even when $\alpha$ is an obtuse angle (i.e., the supplement angle of $\alpha$), the change in pixel scale of the focal plane caused by velocity aberration remains governed by formula (\ref{eq03}). In essence, whether the telescope is oriented toward or moving away from the star, the pixel scale alterations in the focal plane induced by velocity aberration remain consistent. It depends solely on the angle ($\alpha$) between the telescope's motion direction and the line of sight when it maintains a constant speed.

\section{Simulations}
\label{sect:sim}
\subsection{CSST Orbit Simulation}
\label{subsect:data}
Based on theoretical analysis, we calculated the velocity aberration effect, accounting for the orbital motion of CSST within one orbital period and the Earth's annual revolution. Our study employed astrometry-related processing utilizing the Standards Of Fundamental Astronomy (SOFA) library, adhering to the IAU official specifications (IERS 2021). The model's precision is capable of reaching a level of ten milliarcseconds.

\begin{figure}
   \centering
   \includegraphics[width=\textwidth, angle=0]{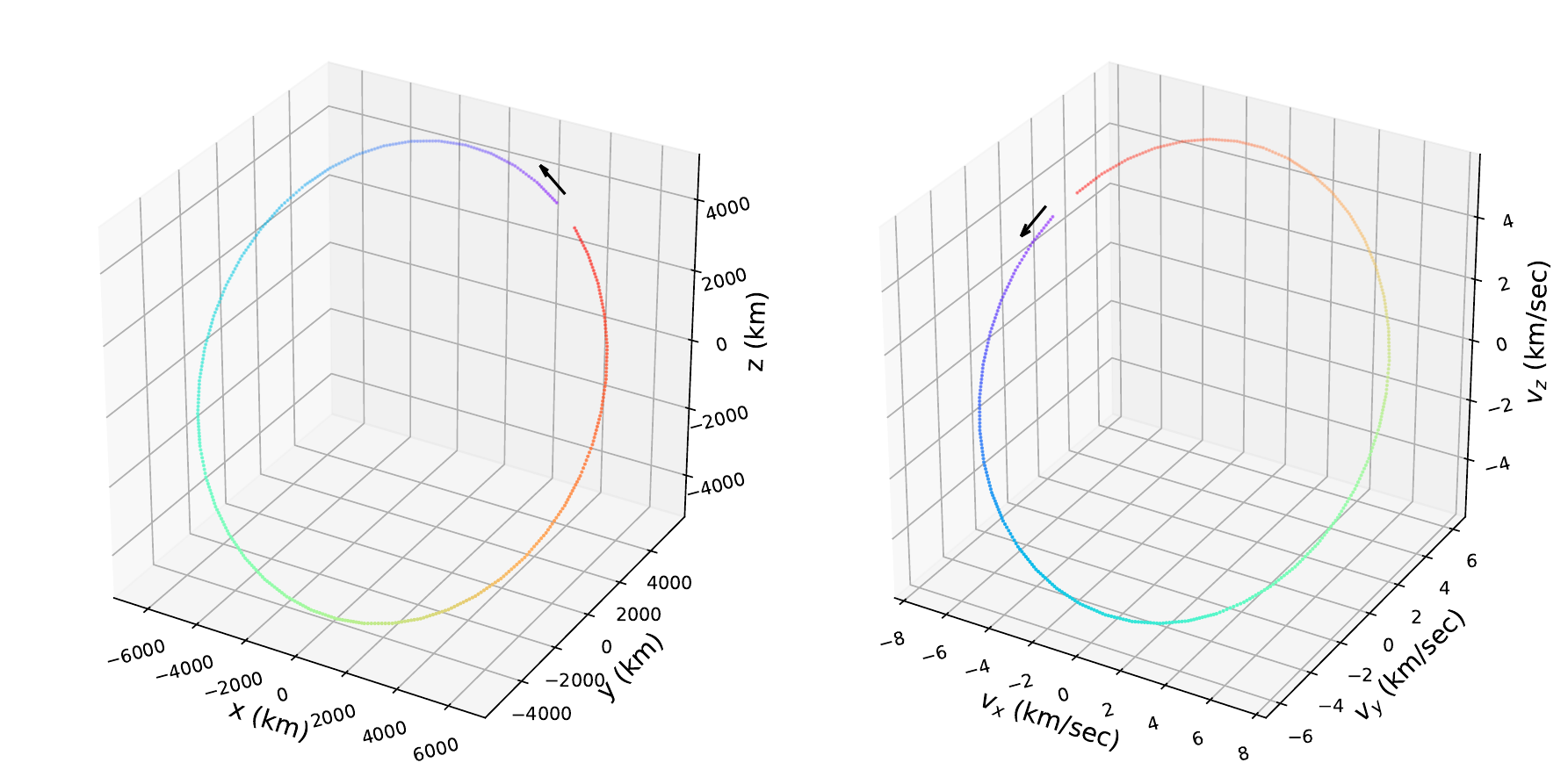}
   \caption{Distribution of orbital position ($left$) and velocity ($right$) in one orbit. The rainbow colors represent the time of one orbit. The black arrows indicate the ascending direction of the time.}
   \label{Fig02}
   \end{figure}

We utilized simulated orbit data for CSST to facilitate this analysis. Initially, we obtained concise orbital data covering a relatively extended period. For the one-orbit total velocity aberration effect analysis, we specifically selected orbit data encompassing a single orbital period (approximately 90 minutes), ranging from "2022-07-06 01:33:40" to "2022-07-06 03:03:40" in UTC. The orbital data within this duration is about a 2-minute interval between consecutive data points. Figure \ref{Fig02} illustrates the orbital position (left) and velocity (right) during this period. The color gradient represents time progression, and the black arrow indicates the direction of CSST motion along the orbit. For the annual velocity aberration effect analysis, we set the orbital position and velocity to zero, with a time interval of one hour between data points.

\subsection{Simulation of the velocity aberration for the CSST MSC }
\label{subsect:method}
\begin{figure}
   \centering
   \includegraphics[width=1.0\textwidth, angle=0]{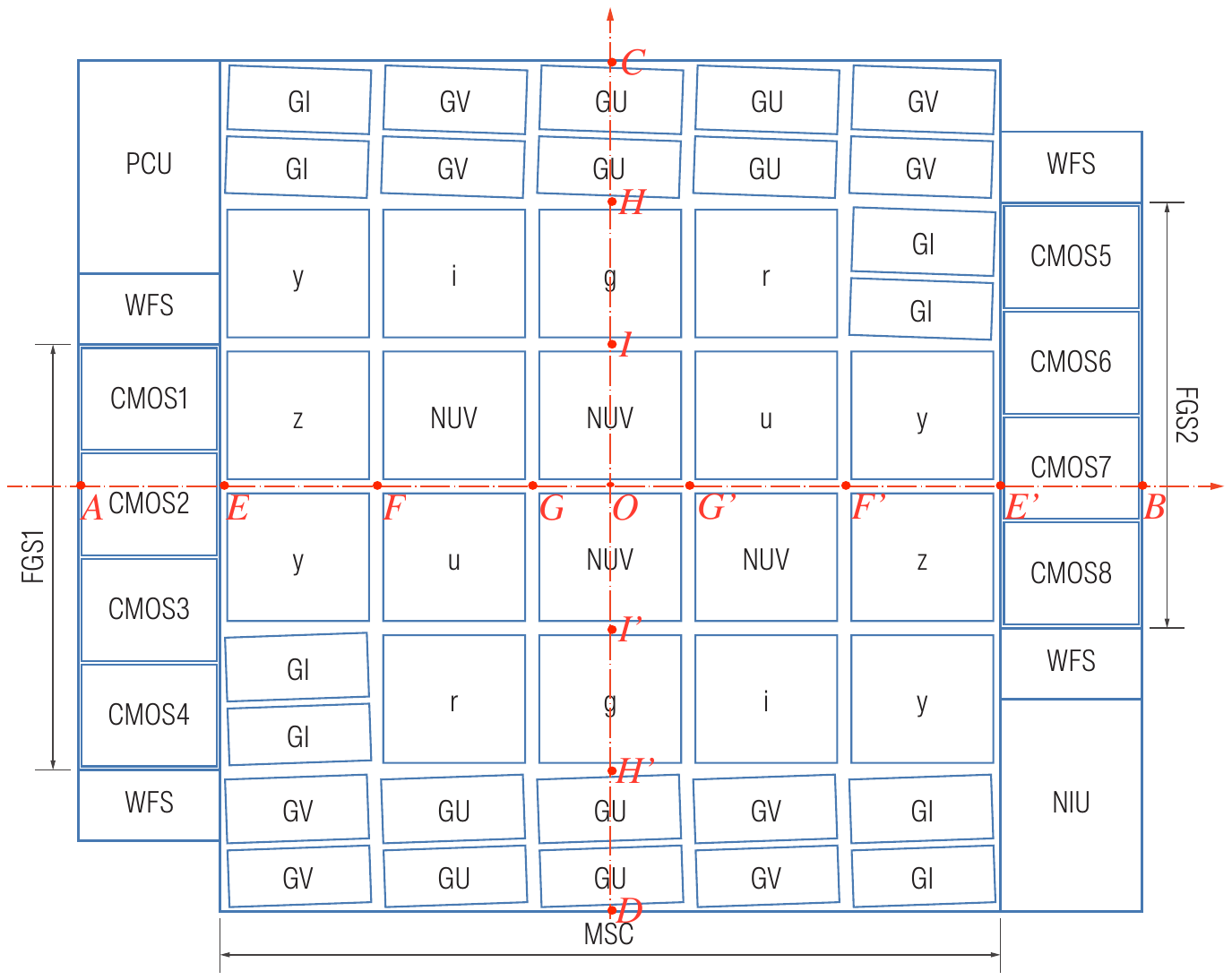}
   \caption{The locations of reference points on the CSST MSC. The MSC comprises 30 detectors, with 18 dedicated to multi-band imaging observations (NUV, $u$, $g$, $r$, $i$, $z$, $y$), and the remaining 12 are used for spectroscopic observations (GU, GV, GI). Within the slightly lower image quality fields of view on both sides, the MSC incorporates the Photometric Calibration Unit (PCU), Near-infrared Unit (NIU), Wavefront Sensor (WFS), and Fine Guidance Sensor (FGS).}
   \label{Fig03}
   \end{figure}

This study assumes that variations in pixel scale can adequately represent the velocity aberration effect on CSST MSC. With this assumption as the foundation, our objective is to obtain the distribution of pixel scale changes across CSST MSC as it observes specific sky regions from different orbital positions and with varying speed vectors based on simulation. Assume that the optical axis is perpendicular to the center of the focal plane. We have selected seven pairs of sampling points based on the arrangement of sensors along the symmetric axis of MSC. We have subsequently established seven measurement baselines, denoted as $AB$, $EE^{'}$, $FF^{'}$, $GG^{'}$, $CD$, $HH^{'}$, and $II^{'}$, to represent the pixel scales in general. As illustrated in Figure \ref{Fig03}, a standard coordinate system ($\xi$, $\eta$) has been established on the MSC, and these sampling points have been marked. Table \ref{Tab1} lists these points with their corresponding standard coordinates. The simulation methods are described in detail as follows: 
\begin{itemize}
\item[1.] 
We initiate the calculation of equatorial coordinates for these reference points based on their standard coordinates in the following steps. Given that the CSST sky survey operates in the ecliptic coordinate, we select a specific pointing ($A, D$) in the ecliptic coordinate as the center of the FOV. We then establish standard coordinates ($\xi, \eta$) with the following orientation: the X-axis parallels the ecliptic latitude and follows the direction of increasing longitude, while the Y-axis is perpendicular to the X-axis and follows the direction of increasing latitude. Subsequently, we employ the Inverse Gnomonic Projection, as indicated in Equation (\ref{eq1}) and (\ref{eq2}) (\citealt{Dick1991}), to transform the standard coordinates ($\xi, \eta$) of these reference points into ecliptic coordinates ($\lambda, \beta$). These ecliptic coordinates are then further converted into equatorial coordinates.
\begin{equation}\label{eq1}
   \lambda = \arctan(\frac{\xi}{\cos{D} - \eta \sin{D}}) + A
\end{equation}
\begin{equation}\label{eq2}
   \beta = \arctan(\frac{\eta \cos{D} + \sin{D}}{\cos{D} - \eta \sin{D}} \cos{(\lambda - A)})
\end{equation}

\item[2.] We compute the equatorial coordinates for the reference points, accounting for both the presence and absence of velocity aberration effects. These reference points, represented by their equatorial coordinates, can be regarded as celestial objects. Several astrometric effects should be involved when considering the light path from these celestial objects in the CSST focal plane. Our analysis concentrates solely on the velocity aberration effects while disregarding other factors, such as projection effects, coordinate transformation errors, etc.
Utilizing the simulated orbit data as input, we employ the Control Variable Method to quantify the distinctions between scenarios with and without these velocity aberrations.

\item[3.] We determine the spherical arc length of the equatorial coordinate system for the pairs $AB$, $EE^{'}$, $FF^{'}$, $GG^{'}$, $CD$, $HH^{'}$, and $II^{'}$, by employing a spherical arc length calculation formula, as expressed in Equation (\ref{eq3}).
\begin{equation}\label{eq3} 
   S = R~\arccos[\cos{\beta_1}~\cos{\beta_2}~\cos({\lambda_1} - {\lambda_2}) + \sin{\beta_1}~\sin{\beta_2}]
\end{equation} 
where $R$ represents the radius of the celestial sphere, which we have normalized to a unit value. The pairs of coordinates $(\lambda_1, \beta_1)$ and $(\lambda_2, \beta_2)$, exemplified here with the arc length $AB$, correspond to the equatorial coordinates of the reference points $A$ and $B$.

\item[4.] We deduce the pixel scale change induced by the total or annual velocity aberration. The ratio of arc length with or without velocity aberration indicates the pixel scale change, as defined by Equation (\ref{eq4}).
\begin{equation}\label{eq4}
   Ratio_{arc\;length} = 1 - \cfrac{{arc\;length}_{with\,aberration}}{{arc\;length}_{without\,aberration}}
\end{equation} 

\end{itemize}

\begin{table}
\bc
\begin{minipage}[]{\textwidth}
\caption[]{Standard Coordinates of 7 pairs of reference points.
\label{Tab1}} \end{minipage} \\
\setlength{\tabcolsep}{2mm}
\small
 \begin{tabular}{cccccccc}
  \hline
  \hline\noalign{\smallskip}
\multicolumn{2}{c}{1st pair} & \multicolumn{2}{c}{2nd pair} & \multicolumn{2}{c}{3rd pair} &\multicolumn{2}{c}{4th pair} \\
  \hline\noalign{\smallskip}
$A$ & $B$ & $E$ & $E^{'}$ & $F$ & $F^{'}$ & $G$ & $G^{'}$   \\
(-0.75$^{\circ}$, 0$^{\circ}$) & (0.75$^{\circ}$, 0$^{\circ}$) & (-0.55$^{\circ}$, 0$^{\circ}$) & (0.55$^{\circ}$, 0$^{\circ}$) & (-0.33$^{\circ}$, 0$^{\circ}$) & (0.33$^{\circ}$, 0$^{\circ}$) & (-0.11$^{\circ}$, 0$^{\circ}$) & (0.11$^{\circ}$, 0$^{\circ}$)  \\ \hline
\hline\noalign{\smallskip}
\multicolumn{2}{c}{5th pair} &\multicolumn{2}{c}{6th pair} &\multicolumn{2}{c}{7th pair} & \\ 
  \hline\noalign{\smallskip}
$C$ & $D$ & $H$ & $H^{'}$ & $I$ & $I^{'}$  &\\
(0$^{\circ}$, 0.6$^{\circ}$) & (0$^{\circ}$, -0.6$^{\circ}$) & (0$^{\circ}$, 0.4$^{\circ}$) & (0$^{\circ}$, -0.4$^{\circ}$) & (0$^{\circ}$, 0.2$^{\circ}$) & (0$^{\circ}$, -0.2$^{\circ}$) & \\
  \noalign{\smallskip}\hline
  \hline
\end{tabular}
\ec
\end{table}

\section{Results}
\label{sect:result}

\begin{figure}
\centering
\includegraphics[width=1.0\textwidth, angle=0]{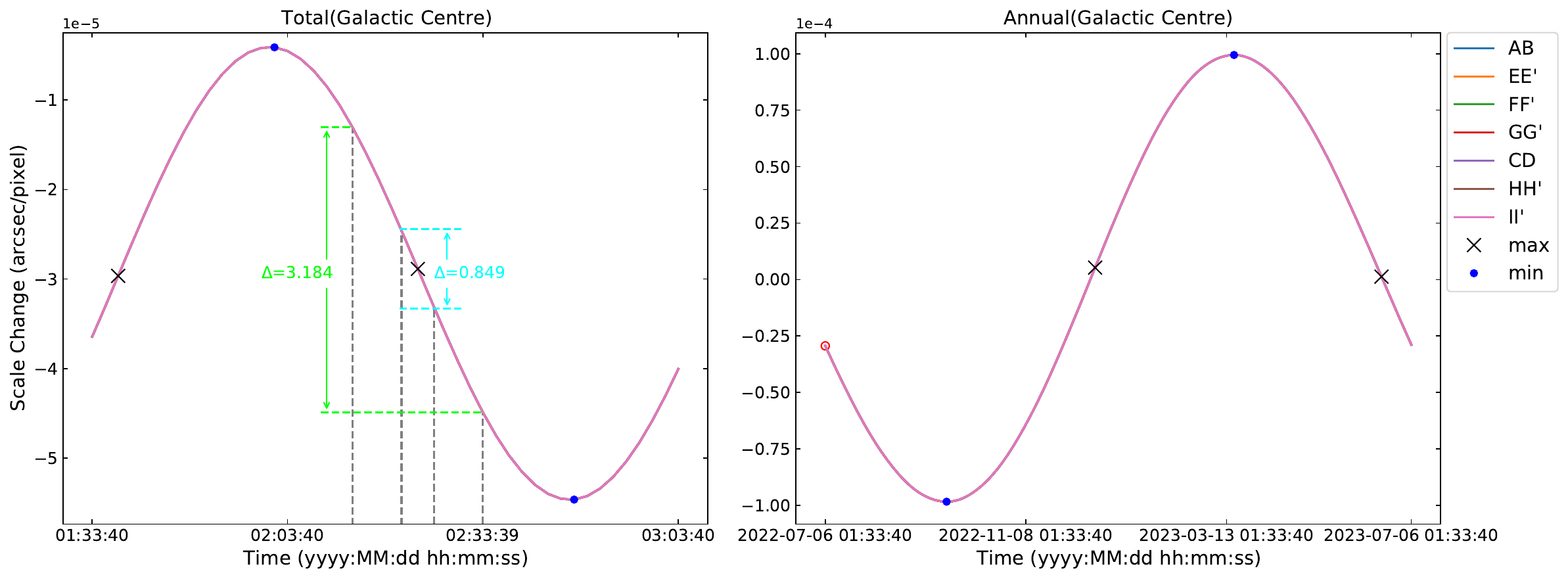}
\caption{
The MSC scale changes result from total velocity aberration (on the left) during one orbit of CSST (about 90 minutes) and annual velocity aberration (on the right) during one year, observed in the direction of the Galactic Center. The horizontal axis on the left represents the time for one orbit, while the right axis corresponds to one year. The lines with different colors, such as AB, EE', FF', GG', CD, HH', and II', represent the scale changes from different measurement baselines, which are overlapped with each other. The maximum and minimum values of the focal plane scale change rate due to total and annual velocity aberration are denoted by black crosses and blue dots, respectively. We use light green to highlight the typical 20-minute long exposure as the maximum values as the center to estimate the maximum total velocity aberration. We use light blue to highlight the typical 300-second ordinary sky survey exposure, the same as the light green. The red circle in the right figure highlights the observation period corresponding to an orbit in the left figure. 
}
\label{galactic_centre}
\end{figure}

We initiated total velocity aberration simulations during one orbit of CSST when the CSST is directed toward the galactic center to investigate the pixel scale variations. The results of this simulation are illustrated in Figure \ref{galactic_centre}. The left subgraph of Figure \ref{galactic_centre} illustrates the simulated variation in the scale of the MSC caused by the total velocity aberration during one orbit of the CSST, which lasts approximately 90 minutes when observed in the direction of the Galactic Center. In the figure, we have denoted the maximum and minimum gradients with black crosses and blue points, respectively. To estimate the maximum total velocity aberration, we have highlighted the CSST typical long exposure (20 minutes) with light green as the black cross in the center. Similarly, we have used light blue to emphasize the typical 300-second exposure for CSST ordinary sky surveys. The maximum values for scale change are $3.184\times10^{-5}$ arcsec/pixel and $0.849\times10^{-5}$ arcsec/pixel for velocity aberration in 20-min and 300-s, respectively. Assuming a Field of View (FOV) of 1.2 degrees, this pixel scale change of the long exposure approximately amounts to 0.069 arcseconds. Consequently, during 20-min exposure, a star image located at the edge of the focal plane undergoes an approximate shift of 0.929 pixels as the plate scale value of the MSC is 0.074 arcsec/pixel. Additionally, we had annual velocity aberration simulations for one year in the same sky area. As the right subgraph of Figure \ref{galactic_centre} shows, the changes of different measurement baselines on MSC coincide, indicating these pixel scale change rates and trends are the same.

\begin{figure}
\centering
\includegraphics[width=1.0\textwidth, angle=0]{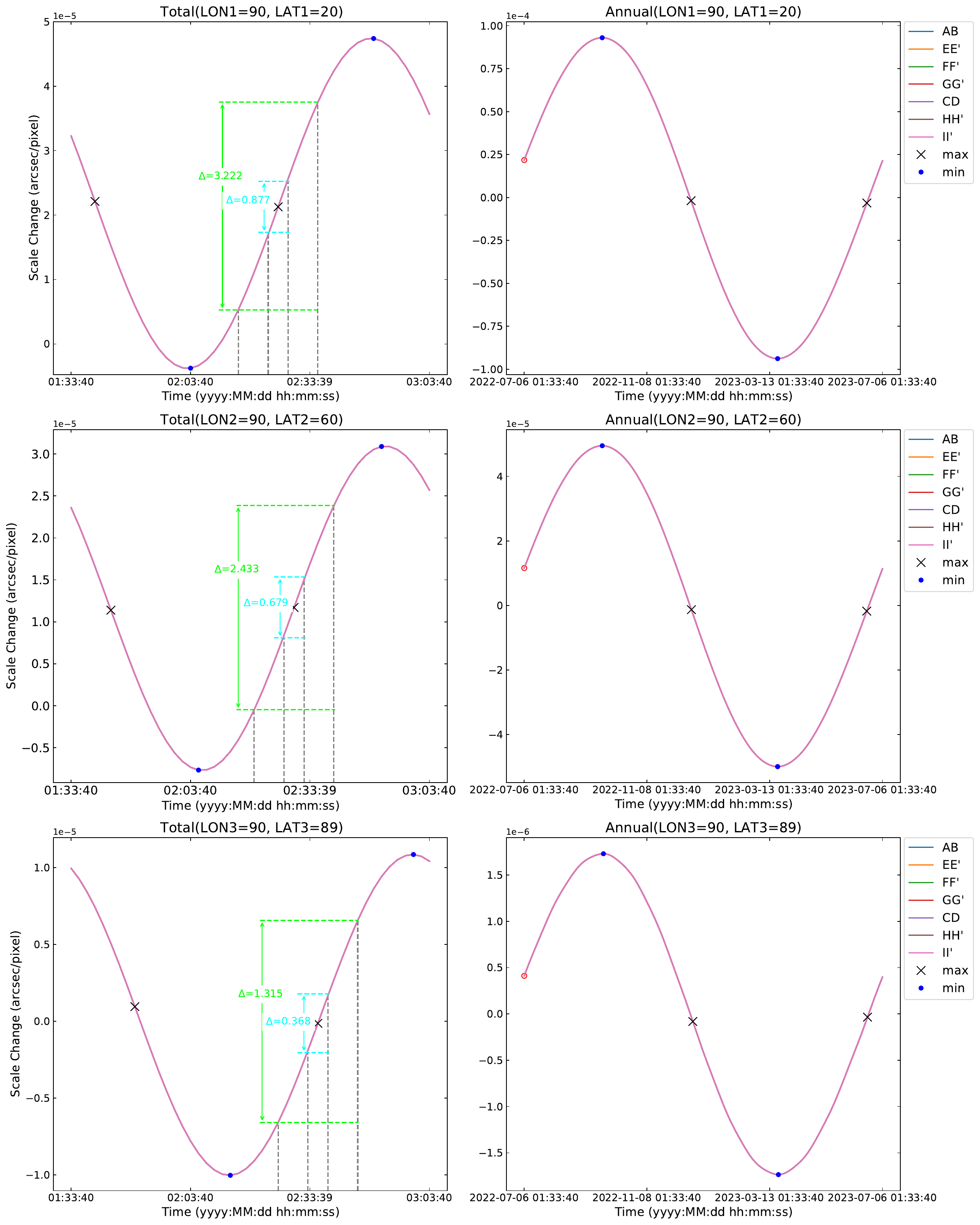}
\caption{
The MSC scale changes result from total velocity aberration (on the left) during one orbit of CSST (about 90 minutes) and annual velocity aberration (on the right) during one year, observed in the directions of the different ecliptic positions. The ecliptic longitude remains fixed at 90 degrees, while the ecliptic latitudes vary between 20$^{\circ}$, 60$^{\circ}$, and 89$^{\circ}$. The horizontal axis on the left represents the time for one orbit, while the right axis corresponds to one year. LON and LAT stand for ecliptic longitude and latitude, respectively. Different colors, such as AB, EE', FF', GG', CD, HH', and II', represent scale changes from different baselines, which overlap. The maximum and minimum values of the focal plane scale change rate due to total and annual velocity aberration are denoted by black crosses and blue dots, respectively. We use light green to highlight the typical 20-minute long exposure as the maximum values at the center to estimate the maximum total velocity aberration. We use light blue to highlight the typical 300-second ordinary sky survey exposure, the same as the light green. The red circle in the right figure highlights the observation period corresponding to an orbit in the left figure. 
}
\label{Fig04}
\end{figure}

We conducted similar simulations to assess both the total velocity aberration effect over a single CSST orbit and the annual velocity aberration effect over one year when CSST observed at different ecliptic latitudes (20$^{\circ}$, 60$^{\circ}$, 89$^{\circ}$). The outcomes of these simulations are presented in Figure \ref{Fig04}. In the left column of the subgraphs, we observe the variations in the total velocity aberration effects on the MSC during one CSST orbit, showcasing different trends. Notably, the discrepancy between the maximum and minimum values of pixel scale changes decreases as the ecliptic latitude rises. Meanwhile, the right column of the subgraphs illustrates the impact of annual velocity aberration. Remarkably, the trends across different measurement baselines within the MSC align, indicating that these rates of scale change remain consistent. Furthermore, we observe that akin to the total velocity aberration, the difference between the maximum and minimum values of scale changes diminishes with higher ecliptic latitudes.

As presented in Table \ref{Tab2}, we have compiled the maximum variations in focal plane scale for 20-minute and 300-second exposures in the context of the three selected sky regions also presented in Figure \ref{Fig04}. The second and third columns provide the FOV centers in ecliptic coordinates ($\lambda, \beta$). The fourth column lists the maximum pixel scale change caused by total velocity aberration, corresponding to the positions marked by black crosses in the left sub-graph of Figure \ref{Fig04}. The fifth column presents the positional change for the edge of FOV, measured in arcseconds. The last column displays the corresponding maximum offset in pixels, offering a more intuitive and direct representation.

\begin{figure}
\centering
\includegraphics[width=1.0\textwidth, angle=0]{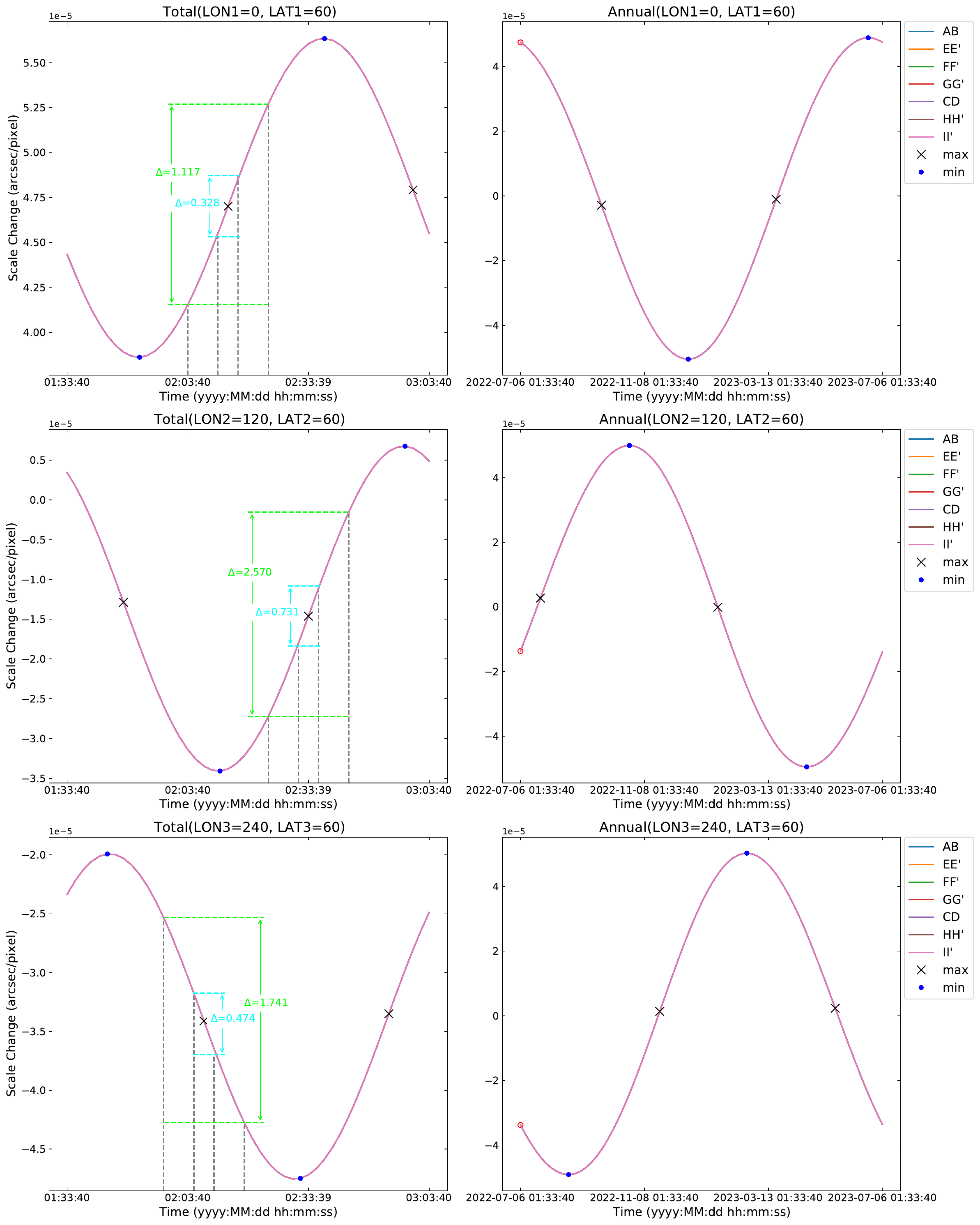}
\caption{Similar to Figure \ref{Fig04}, the MSC scale changes result from total velocity aberration (on the left) and annual velocity aberration (on the right) observed at different ecliptic positions. The ecliptic latitude is fixed at 60$^{\circ}$, while ecliptic longitudes are 0$^{\circ}$, 120$^{\circ}$, 240$^{\circ}$, respectively. }
\label{Fig05}
\end{figure}

Furthermore, we conducted tests to assess the changes in pixel scale under varying longitudes while maintaining the same latitudes. The results, as depicted in Figure \ref{Fig05}, reveal a sinusoidal distribution trend akin to that seen in Figure \ref{Fig04}. However, unlike the trend observed with increasing ecliptic latitudes, the difference between the maximum and minimum values of the pixel scale change remains consistent. Therefore, it becomes apparent that the effect of velocity aberration on the focal plane scale is more pronounced when the observed object is in proximity to the orbital plane of the telescope. In contrast, the impact is independent of changes in ecliptic longitudes.

\begin{figure}
\centering
\includegraphics[width=1.0\textwidth, angle=0]{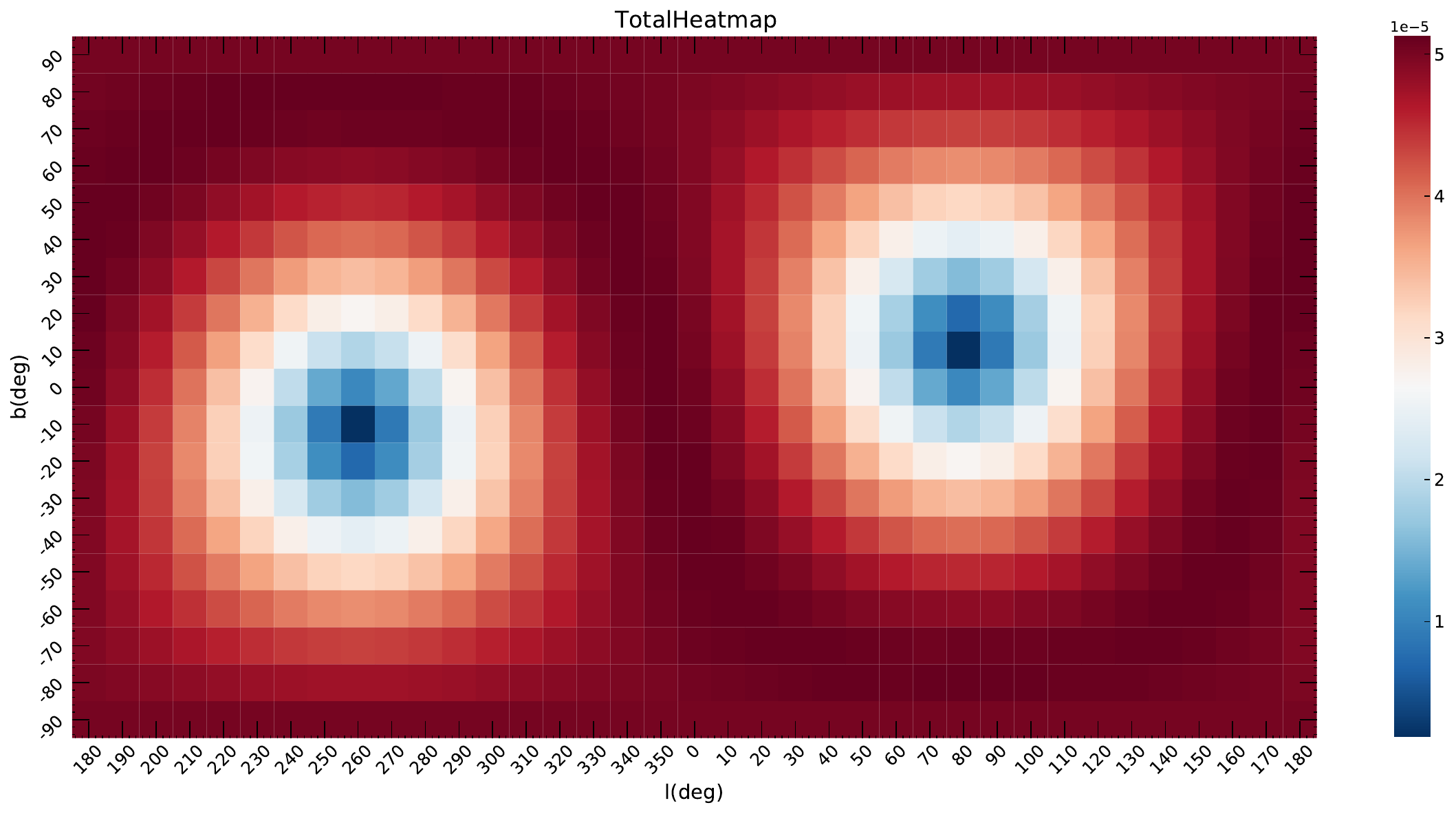}
\caption{The difference between the maximum and minimum values of the MSC pixel scale change caused by the total velocity aberration in one orbit varies with the spatial distribution (Galactic coordinate system, with the Galactic Center at the center).}
\label{Fig07}
\end{figure}

\begin{figure}
\centering
\includegraphics[width=1.0\textwidth, angle=0]{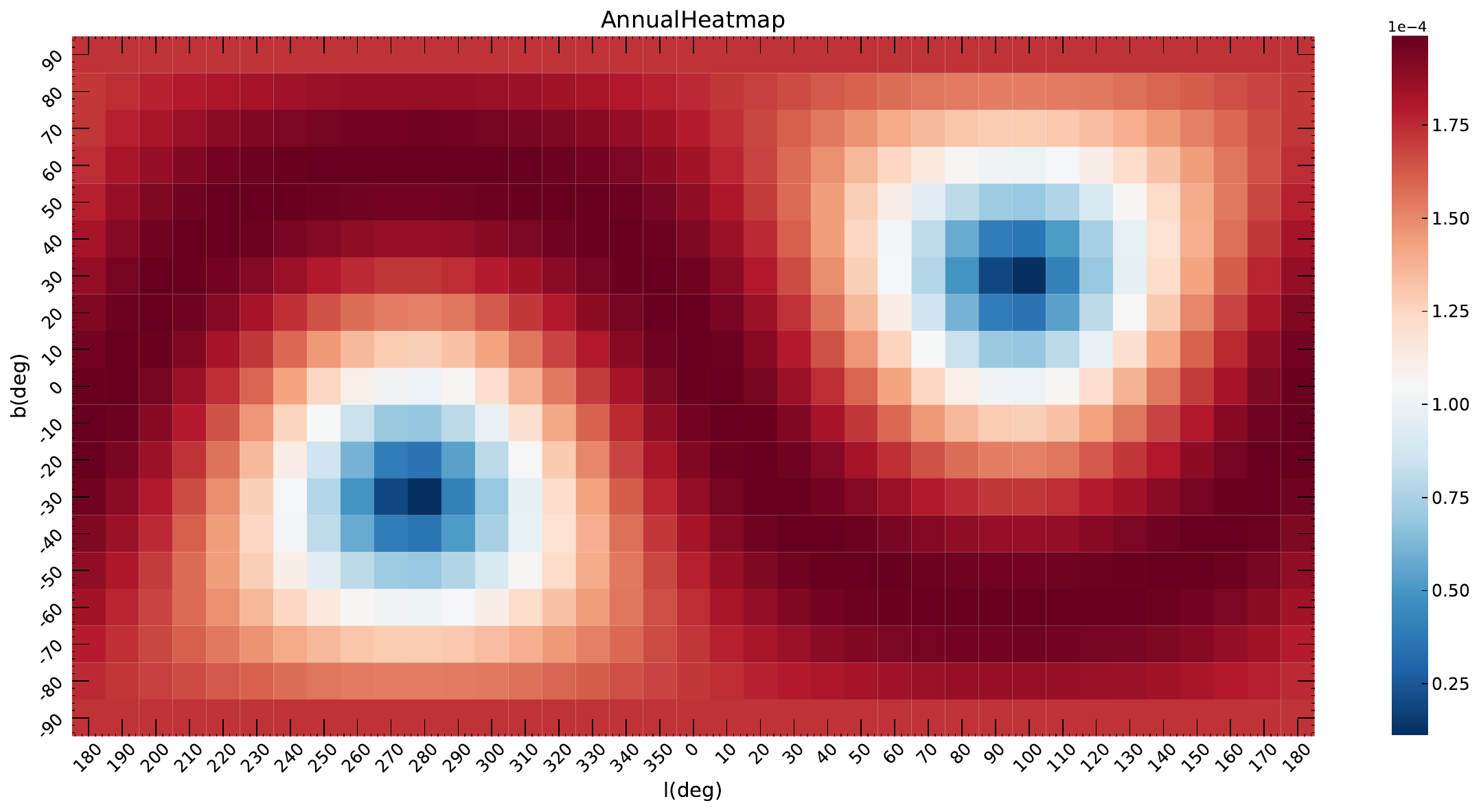}
\caption{The difference between the maximum and minimum values of the MSC pixel scale change caused by the annual velocity aberration in one year varies with the spatial distribution (Galactic coordinate system, with the Galactic Center at the center).}
\label{Fig08}
\end{figure}

The Galactic Bulge is a crucial component that preserves the Milky Way galaxy's early evolutionary trajectory (\citealt{2018ARA&A..56..223B}), but more observational evidence is needed. The insufficient observational data is primarily due to the high extinction and stellar density in the Milky Way's central region, making it difficult to achieve the required observational depth and precise positional measurements of individual stars. The high image quality and resolution of CSST could provide an opportunity to obtain sufficient observations in different epochs. As mentioned above, we should estimate the velocity aberration effect following the galactic coordinate system. Figure \ref{Fig07} and Figure \ref{Fig08} visually depict the disparity between the maximum and minimum values of the MSC scale change caused by total and annual velocity aberration, respectively, as functions of Galactic coordinates. The Galactic Center is centrally positioned within both images. These figures exhibit strikingly similar patterns. It is worth noting that the velocity aberration effect presents a pronounced variation within the nuclear bulge region near the Galactic center, as evidenced by the deep red coloring. Additionally, the effect appears minimal at the north and south Ecliptic poles, as indicated by the deep blue coloring. A more pronounced gradient in the distribution of the velocity aberration effect is observed along the direction of ecliptic latitude compared to the longitude direction.

\begin{table}
\bc
\begin{minipage}[]{\textwidth}
\caption[]{The largest changes were caused by the total velocity aberration in a 20-minute or 300-second exposure of one orbit on the CSST MSC.
\label{Tab2}} \end{minipage} \\
\setlength{\tabcolsep}{3mm}
\small
 \begin{tabular}{cccccc}
  \hline
  \hline\noalign{\smallskip}
\multirow{2}*{Exposure Time} & \multicolumn{2}{c}{Pointing} & Pixel Scale Change& FOV Edge Change & FOV Edge Change\\
  ~ & $\lambda$ & $\beta$   & $\times 10^{-5}$ & arcsec & pixel\\
  \hline\noalign{\smallskip}
\multirow{3}*{20-minute} & 90$^{\circ}$ & 20$^{\circ}$ & 3.222 & 0.070 & 0.940 \\
~ & 90$^{\circ}$ & 60$^{\circ}$ & 2.433 & 0.053 & 0.710 \\
~ & 90$^{\circ}$ & 89$^{\circ}$ & 1.315 & 0.028 & 0.384 \\
\hline
\multirow{3}*{300-second} & 90$^{\circ}$ & 20$^{\circ}$ & 0.877 & 0.019 & 0.256 \\
~ & 90$^{\circ}$ & 60$^{\circ}$ & 0.679 & 0.015 & 0.198 \\
~ & 90$^{\circ}$ & 89$^{\circ}$ & 0.368 & 0.008 & 0.107 \\  \noalign{\smallskip}\hline
  \hline
\end{tabular}
\ec
\end{table}

\section{Conclusions}
\label{sect:conclusion}

The characteristics of the CSST MSC velocity aberration effect, obtained from the theoretical analysis and simulation results of this study, can be summarized as follows:
\begin{itemize}
\item[1.] The velocity aberration effect occurs due to the motion of a space telescope and the limited speed of light. This effect impacts the pixel scale of the focal plane, with a more pronounced impact on space observation platforms equipped with larger focal panels.
\item[2.] The velocity aberration effect affects the pixel scale consistently within the MSC across different measurement baselines and in both the vertical and parallel directions.
\item[3.] The pixel scale variation follows a sinusoidal distribution throughout a single orbit, leading to distinct minimum and maximum points for the velocity aberration effect.
\item[4.] It is observed that the influence of velocity aberration on the pixel scale is more pronounced for targets located at lower ecliptic latitudes. In the central region of the Galaxy, the velocity aberration has a significant impact on the focal plane scale. 
\item[5.] At low ecliptic latitudes, the total velocity aberration causes the pixel scale change of approximately $3\times 10^{-5}$ for a 20-minute exposure and $0.9\times 10^{-5}$ for a 300-second exposure. This results in a position offset at the edge of the FOV of about 0.94 pixels for the 20-minute exposure and 0.26 pixels for the 300-second exposure.
\end{itemize}

In conclusion, the velocity aberration effect substantially influences the precision of celestial object measurements acquired through CSST MSC observations, which presents a significant additional data processing for the CSST high-resolution observations. As a result, meticulous consideration and processing of the velocity aberration effect are necessary for the pre- and post-observation phases. 

Since the magnitude of the velocity aberration effect varies according to its inherent characteristics, it becomes feasible to mitigate the effect by strategically utilizing forecasted orbital parameters and the orientation of the observation target. Careful selection of specific start exposure times is essential when planning CSST observation strategies, particularly in the low ecliptic latitude observation areas. 

Further refinement of post-processing techniques can enhance the accuracy of PSF reconstruction and object positioning. These techniques encompass the compilation of correction tables for ePSF data, and PSF reconstruction through multi-Gaussian (Nie et al. in preparing) fitting with additional velocity aberration effect parameters.

Lastly, the influence of the velocity aberration effect on the MSC Fine Guidance Sensors (FGSs) is relevant, and proper consideration should be given to uploading guide star positions and developing on-orbit guiding programs.

\begin{acknowledgements}
First and foremost, we extend our heartfelt thanks to Dr. Qi Z haoxiang and his team for their invaluable support of this study. We acknowledge the support by National Key R\&D Program of China (No. 2022YFF0503403, 2022YFF0711500), the support of National Nature Science Foundation of China (Nos. 11988101, 12073047, 12273077, 12022306, 12373048, 12263005), the support from the Ministry of Science and Technology of China (Nos. 2020SKA0110100),  the science research grants from the China Manned Space Project (Nos. CMS-CSST-2021-B01, CMS-CSST-2021-A01), CAS Project for Young Scientists in Basic Research (No. YSBR-062), and the support from K.C.Wong Education Foundation. This work is based on the mock data created by the CSST Simulation Team, which is supported by the CSST scientific data processing and analysis system of the China Manned Space Project.

\end{acknowledgements}
 
\bibliographystyle{raa}
\bibliography{ref}

\end{document}